\documentclass{aa}    
\usepackage{txfonts}
\usepackage{graphicx}
\usepackage{natbib}
\bibpunct[, ]{(}{)}{,}{a}{}{,}

\usepackage{color}

\newcommand{\logg}{\rm log~ g}
\newcommand{\Teff}{T_{\rm eff}}
\newcommand{\eps}[1]{\log\varepsilon_{\rm #1}}
\newcommand{\epsa}[1]{\log A_{\rm #1}}

\newcommand{\Eexc}{$E_{\rm exc}$}
\newcommand{\eu}[5]{\mbox{$#1\,^#2{\rm #3}^{#4}_{\rm #5}$}}
\newcommand{\kH}{$S_{\!\rm H}$}    
\newcommand{\kms}{km\,s$^{-1}$}
\newcommand{\Vmic}{V_{\rm mic}}

\begin{document}

\title{Astrophysical tests of atomic data \\ important for stellar Mg abundance determinations}
  
\author{
L. Mashonkina\inst{1,2}}

\offprints{L. Mashonkina; \email{lima@inasan.ru}}
\institute{
     Universit\"ats-Sternwarte M\"unchen, Scheinerstr. 1, D-81679 M\"unchen, 
     Germany \\ \email{lyuda@usm.lmu.de}
\and Institute of Astronomy, Russian Academy of Sciences, RU-119017 Moscow, 
     Russia \\ \email{lima@inasan.ru}
}

\date{Received  / Accepted }

\abstract {Magnesium abundances of cool stars with different metallicities are important for understanding the galactic chemical evolution.}
{This study tests atomic data used in stellar magnesium abundance analyses.}
{We evaluate non-local thermodynamical equilibrium (NLTE) line formation for \ion{Mg}{i} using the most up-to-date theoretical and experimental atomic data available so far and check the Mg abundances from individual lines in the Sun, four well studied A-type stars, and three reference metal-poor stars.}
{With the adopted $gf$-values, NLTE abundances derived from the \ion{Mg}{i} 4703\,\AA, 5528\,\AA, and \ion{Mg}{i}b lines are consistent within 0.05~dex for each A-type star. The same four \ion{Mg}{i} lines in the solar spectrum give consistent NLTE abundances at $\log N_{\rm Mg}/N_{\rm H} = -4.45$, when correcting van der Waals damping constants inferred from the perturbation theory. Inelastic Mg+H collisions as treated by Barklem, Belyaev, Spielfiedel, Guitou, and Feautrier serve as efficient thermalizing process for the statistical equilibrium of \ion{Mg}{i} in the atmospheres of metal-poor stars. The use of Mg+H collision data improves Mg abundance determinations for HD\,84937 and HD\,122563, though does not remove completely the differences between different lines. }
 {}


\titlerunning{Astrophysical tests of atomic data} 

\authorrunning{L. Mashonkina}

\maketitle

\section{Introduction}

Magnesium plays an important role in studies of cool stars. It affects atmospheric structure through donation of free electrons and significant contribution of \ion{Mg}{i} photoionization to the ultra-violet (UV) opacity. Magnesium is a key element for studing the history of $\alpha-$process nucleosynthesis in the Universe. The Mg/Fe abundance ratio of metal-poor (MP) stars carries on an information on the initial mass function and star formation rate in our Galaxy, dwarf galaxies of the Local Group, and distant galaxies \citep[see][for references]{2009ARA&A..47..371T}. Magnesium is one of the best observed elements in A to late type stars, though it is represented by only few lines of neutral atoms in the visual spectral range. The \ion{Mg}{i}b lines at 5167, 5172, and 5183\,\AA\ can be measured over a wide range of metallicity from super-solar values down to [Fe/H]\footnote{In the classical notation, where [X/H] = $\log(N_{\rm
    X}/N_{\rm H})_{star} - \log(N_{\rm X}/N_{\rm H})_{Sun}$.} = $-5.5$ \citep{Frebeletal:2005} and over a wide range of effective temperature from 3000~K up to 12000-13000~K \citep{Przybilla_mg}. 
In stellar atmospheres with $\Teff > 4500$~K, neutral magnesium is a minority species, and its statistical equilibrium (SE) can easily deviate from thermodynamic equilibrium due to deviations of the mean intensity of ionizing radiation from the Planck function. Since the end of the 1960s, the problem of non-local thermodynamic equilibrium (NLTE) line formation for \ion{Mg}{i} in stellar atmospheres was considered in many studies. The model atoms of \ion{Mg}{i} were created by 
\citet{Athay1969mg,Lemke1987,Mauas1988,Gigas88,Carlsson92,1996ARep...40..187M,Zhao1998,Gratton1999,Thevenin2000,Przybilla_mg,Mishenina2004,2011MNRAS.418..863M}.

This study was motivated by recent detailed quantum mechanical calculations of \citet[][hereafter, BBSGF]{mg_hyd2012} for inelastic Mg+H collisions. The role of 
inelastic collisions with neutral hydrogen atoms in establishing the SE of atoms in cool stars is debated for decades. \citet{Steenbock1984} implemented the classical \citet{Drawin1968,Drawin1969} formula to calculate \ion{H}{i} collision rates, and it suggests that their influence is comparable to electron impacts. Till recent time, laboratory measurements and/or detailed quantum mechanical
calculations of collisions with \ion{H}{i} atoms were only available for \ion{Na}{i} and \ion{Li}{i} \citep[see][for references]{Barklem2011_hyd}. Based on these data, the Drawin's formalism has been criticized for 
not providing a realistic description of the physics involved and
 overestimating the collision rates. Interestingly, spectroscopic studies of \ion{Na}{i} also suggest that the Drawin's formula strongly overestimates the Na+H collision rates.
For example, \citet{h_cool_na_o} reproduced the center-to-limb variation of the solar \ion{Na}{i} 6160\AA\ line with pure electronic collisions. 
At the same time, the need
for a thermalizing process not involving electrons in the atmospheres
of, in particular, very metal-poor stars, was indicated by
many NLTE line-formation studies \citep[see][and references therein]{mash_fe}.

The main goal of this paper is to investigate how the use of recent data of BBSGF on Mg+H collisions influences the NLTE results for \ion{Mg}{i}. We started from analysis of the solar \ion{Mg}{i} lines, for which the departures from LTE are small and the derived abundances are not influenced by \ion{H}{i} collision treatment. It was found that the two lines \ion{Mg}{i} 4703\,\AA\ and \ion{Mg}{i} 5528\,\AA\ give significantly lower abundance compared with that from the \ion{Mg}{i}b lines and also compared with the meteoritic value. To understand the source of discrepancies and to separate effects of oscillator strengths and van der Waals broadening, we moved to  
the hotter A-type stars.
The paper is organized as follows. All our results are based on the NLTE line formation for \ion{Mg}{i}.
The model atom of magnesium and the used atomic data are described in Sect.\,\ref{sect:NLTE}. In Sect.\,\ref{sect:sun}, we derive abundances from individual \ion{Mg}{i} lines in the solar spectrum and in the spectra of four well studied A-type stars. Section\,\ref{sect:collisions} investigates the effect of Mg+H collisions on the SE of magnesium and NLTE abundances for the two very metal-poor ([Fe/H] $< -2$, VMP) stars. We compare the NLTE results from calculations using Mg+H collision data of BBSGF and using the classical Drawinian rates. The NLTE abundance corrections are presented for lines of \ion{Mg}{i} in the grid of metal-poor model atmospheres. 
Our conclusions are given in Sect.\,\ref{conclusion}.

\section{Method of NLTE calculations for magnesium}\label{sect:NLTE}

This study uses a comprehensive model atom that includes 85 levels of \ion{Mg}{i}, 2 levels of \ion{Mg}{ii}, and the ground state of \ion{Mg}{iii}. For \ion{Mg}{i}, we rely on the model atom produced by \citet{Zhao1998}, who carefully investigated atomic data on the energy levels, transition probabilities, and photoionization cross sections. In this study, the calculation of collisional rates was updated. For electron-impact excitation, we used rate coefficients of \citet{Mauas1988}, where available, and the same recipes as in \citet{Zhao1998} for the remaining transitions. Ionization by electronic collisions is calculated from the \citet{Seaton1962} formula with a mean Gaunt factor set equal to $\overline{g}$ = 0.1 for \ion{Mg}{i} and to 0.2 for \ion{Mg}{ii}.
 For \ion{H}{i} impact excitations and charge transfer processes \ion{Mg}{i} + \ion{H}{i} $\leftrightarrow$ \ion{Mg}{ii} + H$^-$, rate coefficients were taken from detailed quantum mechanical calculations of BBSGF.

Singly ionized magnesium is represented in our model atom by the two levels \eu{3s}{2}{S}{}{} and \eu{3p}{2}{P}{\circ}{}. Their energies and $gf$-value of the resonance transition were taken from the NIST database\footnote{\tt  http://www.nist.gov/pml/data/asd.cfm} \citep{NIST}. Photoionization is treated by utilizing Opacity Project cross-sections as available through the TOPBASE database\footnote{\tt http://cdsweb.u-strasbg.fr/topbase/xsections.html}.

In order to solve the coupled radiative transfer and statistical
equilibrium equations, we used a revised version of the
DETAIL program \citep{detail}.
The update was described by \citet{2011JPhCS.328a2015P}.

Calculations were performed with plane-parallel (1D), 
LTE, and blanketed model atmospheres computed for given stellar parameters. 
These are the {\sc MAFAGS-OS} models by \citet{Grupp2009} for the Sun and VMP stars, the {\sc LLMODELS} models \citep{LLMODELS} by D.~Shulyak for the three A-type stars, and the model by R.~Kurucz\footnote{\tt http://kurucz.harvard.edu/stars/SIRIUS/} for Sirius.

\begin{table}
 \centering
 \caption{\label{line_data} \ion{Mg}{i} line data.}
  \begin{tabular}{ccccc}
   \hline\noalign{\smallskip}
$\lambda$ & \Eexc & log $gf$ & $\log \Gamma_4/N_{\rm e}$ & $\log \Gamma_6/N_{\rm H}$ \\
(\AA)     & (eV)  &          & (rad/s$\cdot$cm$^3$) & (rad/s$\cdot$cm$^3$)  \\
\noalign{\smallskip} \hline \noalign{\smallskip}
4571.10 & 0.00 & --5.62$^1$ & --6.46 & --7.77  \\
4702.99 & 4.33 & --0.44$^2$ & --4.17 & --6.69  \\
5528.41 & 4.33 & --0.50$^2$ & --4.63 & --6.98  \\
5172.68 & 2.71 & --0.45$^3$ & --5.43 & --7.27  \\
5183.60 & 2.71 & --0.24$^3$ & --5.43 & --7.27   \\
\noalign{\smallskip} \hline \noalign{\smallskip}
\multicolumn{5}{l}{$\Gamma_4$ and $\Gamma_6$ correspond to 10\,000~K. } \\
\multicolumn{5}{l}{Notes. \ $^1$ NIST; \ \ $^2$ \citet{1990JQSRT..43..207C}; \ \ $^3$ \citet{Aldenius_mg1}. } \\
\end{tabular}
\end{table}

Table~\ref{line_data} lists the lines of \ion{Mg}{i} used in abundance analyses for MP stars together with the adopted line data. Oscillator strengths 
were taken from laboratory measurements of \citet{Aldenius_mg1} for the \ion{Mg}{i}b lines, calculations of \citet{1990JQSRT..43..207C} for \ion{Mg}{i} 4703\,\AA\ and 5528\,\AA, and calculations of Tachiev \& Froese Fischer for \ion{Mg}{i} 4571\,\AA\ as presented in the NIST database. 
 An accuracy of predicted $gf$-values can only be estimated for the \ion{Mg}{i}b lines, where the difference between calculations of \citet{1990JQSRT..43..207C} and measurements of \citet{Aldenius_mg1} amounts to 0.06~dex.
The van der Waals broadening constants $\Gamma_6$ were computed with cross-sections and velocity parameters from \citet{1995MNRAS.276..859A} and \citet{1997MNRAS.290..102B}. Hereafter, these papers by  Anstee, Barklem, and 
O'Mara are referred to as the $ABO$ theory.
The quadratic Stark effect broadening constants $\Gamma_4$ were taken from \citet{mg1_g4}, except for \ion{Mg}{i} 4571\,\AA, with $\Gamma_4$ from the VALD database \citep{vald}.

\begin{table*}
 \centering
 \caption{\label{mg_eps1} Magnesium NLTE and LTE abundances $\epsa{Mg}$ of the Sun and hot stars.}
  \begin{tabular}{cccccccrcccrcccrcc}
   \hline\noalign{\smallskip}
$\lambda$ & \multicolumn{2}{c}{Sun} & & \multicolumn{2}{c}{HD\,32115} & &\multicolumn{3}{c}{Vega} & &\multicolumn{3}{c}{Sirius} & &\multicolumn{3}{c}{21 Peg} \\
\noalign{\smallskip} 
\cline{2-3} 
\cline{5-6} 
\cline{8-10} 
\cline{12-14}
\cline{16-18} 
(\AA)     & NLTE & LTE & & NLTE & LTE & & $EW$$^1$ & NLTE & LTE & & $EW$ & NLTE & LTE & & $EW$ & NLTE & LTE \\
\noalign{\smallskip} \hline \noalign{\smallskip}
\ion{Mg}{i} 4571 & --4.38 & --4.42 & & -      & -      & & -   & -	& -	 & & -   & -	  & -	   & & -  & -	   & - \\
\ion{Mg}{i} 4703 & --4.69 & --4.70 & & --4.55 & --4.58 & &  31 & --4.93 & --4.94 & &  42 & --4.56 & --4.56 & & 17 & --4.48 & --4.50 \\
\ion{Mg}{i} 5528 & --4.60 & --4.61 & & --4.52 & --4.54 & &  27 & --4.93 & --4.95 & &  39 & --4.54 & --4.53 & & 14 & --4.49 & --4.51 \\
\ion{Mg}{i} 5172 & --4.45 & --4.46 & & --4.44 & --4.44 & & 106 & --4.88 & --4.74 & & 121 & --4.44 & --4.25 & & 70 & --4.40 & --4.30 \\
\ion{Mg}{i} 5183 & --4.45 & --4.46 & & --4.44 & --4.45 & & 124 & --4.84 & --4.62 & & 134 & --4.44 & --4.21 & & 86 & --4.35 & --4.16  \\
\noalign{\smallskip} \hline \noalign{\smallskip}
 & \multicolumn{17}{c}{ \ \ \ $\Delta\epsa{}$(\ion{Mg}{i} 4703,5528 - \ion{Mg}{i} 5172,5183)} \\
\noalign{\smallskip} \hline \noalign{\smallskip}
                 & --0.19 &        & & --0.10 &        & &     & --0.07 &        & &     & --0.11 &        & &    & --0.11 &  \\
\noalign{\smallskip} \hline \noalign{\smallskip}
\multicolumn{18}{l}{Note. \ $ ^1$ Equivalent width, $EW$, in m\AA.} \\
\end{tabular}
\end{table*}

\section{\ion{Mg}{i} lines in the Sun and A-type stars}\label{sect:sun}

In the galactic chemical evolution studies, stellar sample covers, as a rule, a range of metallicities of more than 3~dex. The lines \ion{Mg}{i} 4703 and 5528\,\AA\ are favorites for abundance determinations of mildly MP ([Fe/H] $> -2$) stars, however, these lines cannot be measured in hyper metal-poor stars, with [Fe/H] $< -4.5$, where one relies on the \ion{Mg}{i}b lines. One needs, therefore, to prove 
that the use of different \ion{Mg}{i} lines does not produce a systematic shift in derived abundance between various metallicity stars. In the beginning, we checked whether the lines listed in Table~\ref{line_data} give consistent abundances for the Sun, when employing the most up-to-date theoretical and experimental atomic data available so far. Observed solar spectrum was taken from the Kitt Peak Solar Flux Atlas \citep{Atlas}. We used the {\sc MAFAGS-OS} solar model atmosphere with $\Teff$/$\logg$/[Fe/H] = 5780/4.44/0. Microturbulence velocity was fixed at $\Vmic$ = 0.9~\kms. Element abundances from individual lines were determined from line-profile fitting using the code {\sc SIU} \citep{Reetz}. 
The obtained results are presented in Table\,\ref{mg_eps1}. 
We find that (i) the NLTE effects are minor for each investigated line, (ii) abundances derived from the \ion{Mg}{i}b and \ion{Mg}{i} 4571\,\AA\ lines are consistent with the meteoritic value $\epsa{Mg,met}$\footnote{$\epsa{A} = \log N_{\rm A}/N_{\rm H}$.} $= -4.45$ \citep{Lodders2009}, (iii) \ion{Mg}{i} 4703 and 5528\,\AA\ give significantly lower abundances, by 0.24~dex and 0.15~dex, respectively. These discrepancies can arise from the uncertainty in either $gf$-values or/and $\Gamma_6$ values. The influence of atmospheric inhomogeneities on derived solar Mg abundance was evaluated by \citet{Asplund2005ARAA} as $-0.03$~dex.


\begin{table}
  \begin{center}
\caption{Stellar parameters of selected stars.}
\label{startab}
\begin{tabular}{lccrcl}
\noalign{\smallskip} \hline \noalign{\smallskip}
 Object & $\Teff$, K & log\,g & [Fe/H] & $\Vmic ^1$ & Ref. \\
\noalign{\smallskip} \hline 
Sun        & 5780 & 4.44 & 0      & 0.9 & \\
HD\,32115 &  7250 & 4.20 & 0 & 2.3 & F11 \\
HD\,48915 (Sirius) & 9850 & 4.30 & 0.4 & 1.8 & S13 \\
HD\,84937  & 6350 & 4.09 & --2.08 & 1.7 & M11 \\
HD\,103095 & 5070 & 4.69 & --1.35 & 0.8 & M07 \\
HD\,122563 & 4600 & 1.60 & --2.56 & 1.95& M11 \\
HD\,172167 (Vega) & 9550 & 3.95 & --0.5 & 2.0 & P01 \\
HD\,209459 (21 Peg) & 10400 & 3.55 & 0.0 & 0.5 & F09 \\
\noalign{\smallskip} \hline \noalign{\smallskip}
\multicolumn{6}{l}{Note. \ $ ^1$ Microturbulence velocity, in \kms.} \\
\multicolumn{6}{l}{Ref.: F09 = \citet{2009A&A...503..945F}; \ F11 = \citet{2011MNRAS.417..495F};} \\ 
\multicolumn{6}{l}{M07 = \citet{mash_ca}; \ M11 = \citet{mash_fe}; } \\ 
\multicolumn{6}{l}{P01 = \citet{Przybilla_mg}; \  S13 =  \citet{sitnova_o}.} \\
\end{tabular}
 \end{center}
\end{table}

To test $gf$-values, we choose the three well studied stars Sirius, Vega, and 21~Peg (Table\,\ref{startab}),
 where the \ion{Mg}{i} lines are insensitive to $\Gamma_6$ variation. 
 The departures from LTE are expected to be larger in hot atmospheres than in the solar one. Thus, analysis of the \ion{Mg}{i} lines in the A-type stars tests also our NLTE method.
For each star, its stellar parameters and observed equivalent widths indicated in Table\,\ref{mg_eps1} and used to determine element abundances were taken from a common source cited in Table\,\ref{startab}. 
We find that, for each star, (i) the NLTE effects are minor for \ion{Mg}{i} 4703\,\AA\ and 5528\,\AA, (ii) abundances from these two lines are consistent within 0.02~dex, in contrast to the solar case, (iii) departures from LTE for the \ion{Mg}{i}b lines are significant, so that the NLTE abundance correction $\Delta_{\rm NLTE} = \eps{NLTE} - \eps{LTE}$ ranges between $-0.14$~dex and $-0.22$~dex for different stars, (iv) the difference in NLTE abundance between the \ion{Mg}{i}b and \ion{Mg}{i} 4703, 5528\,\AA\ lines is smaller compared to that for the Sun. It is worth noting that it does not exceed 0.05~dex, when applying $gf$-values from a single source \citep{1990JQSRT..43..207C} for all the investigated lines.

Return to the Sun. The abundance discrepancies found between different solar lines of \ion{Mg}{i} can only be owing to 
van der Waals damping treatment, and they are removed by reducing the $\Gamma_6$($ABO$) values by 0.3~dex and 0.2~dex for \ion{Mg}{i} 4703\,\AA\ and \ion{Mg}{i} 5528\,\AA, respectively. 
The use of such corrections was justified by analysis of the \ion{Mg}{i} lines in the cool subdwarf star HD\,103095 (Table~\ref{startab}).
We obtained the difference (\ion{Mg}{i} 4703, 5528 - \ion{Mg}{i}b) = $-0.18$~dex, when using $\Gamma_6$($ABO$), and a significantly smaller value of $+0.02$~dex, with the above corrections. Table\,\ref{mg_eps1} presents Mg abundances of the late A-type star HD\,32115, where \ion{Mg}{i} 4703\,\AA\ and 5528\,\AA\ are less sensitive to $\Gamma_6$ variation. Both lines give identical abundances at $\epsa{Mg} = -4.51$, and the difference (\ion{Mg}{i} 4703, 5528 - \ion{Mg}{i}b) reduces to $-0.07$~dex, when employing the corrected $\Gamma_6$-values. 

\section{Influence of Mg+H collisions on the NLTE results for metal-poor stars}\label{sect:collisions}

\begin{figure}
\flushleft 
  \resizebox{88mm}{!}{\includegraphics{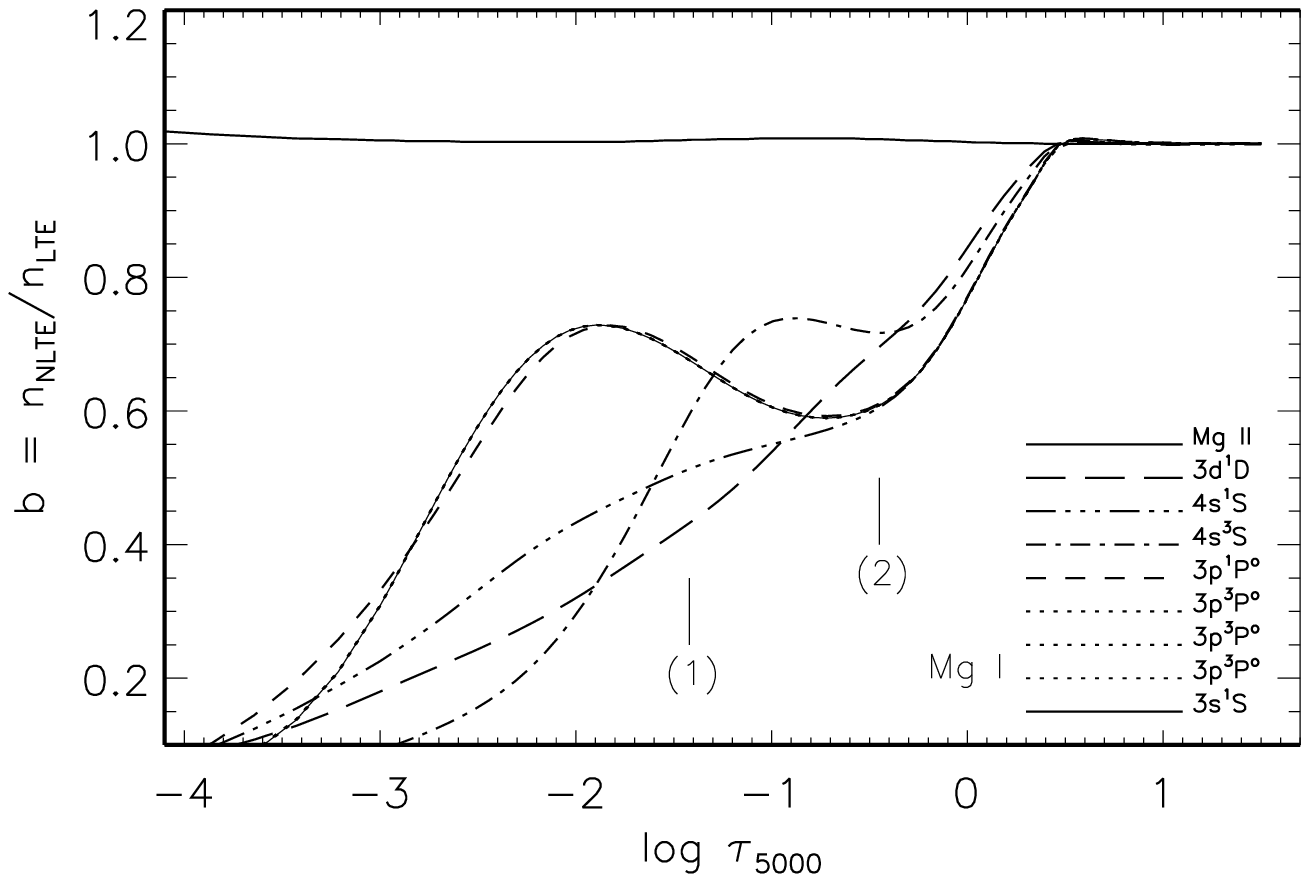}}

\flushleft 
  \resizebox{88mm}{!}{\includegraphics{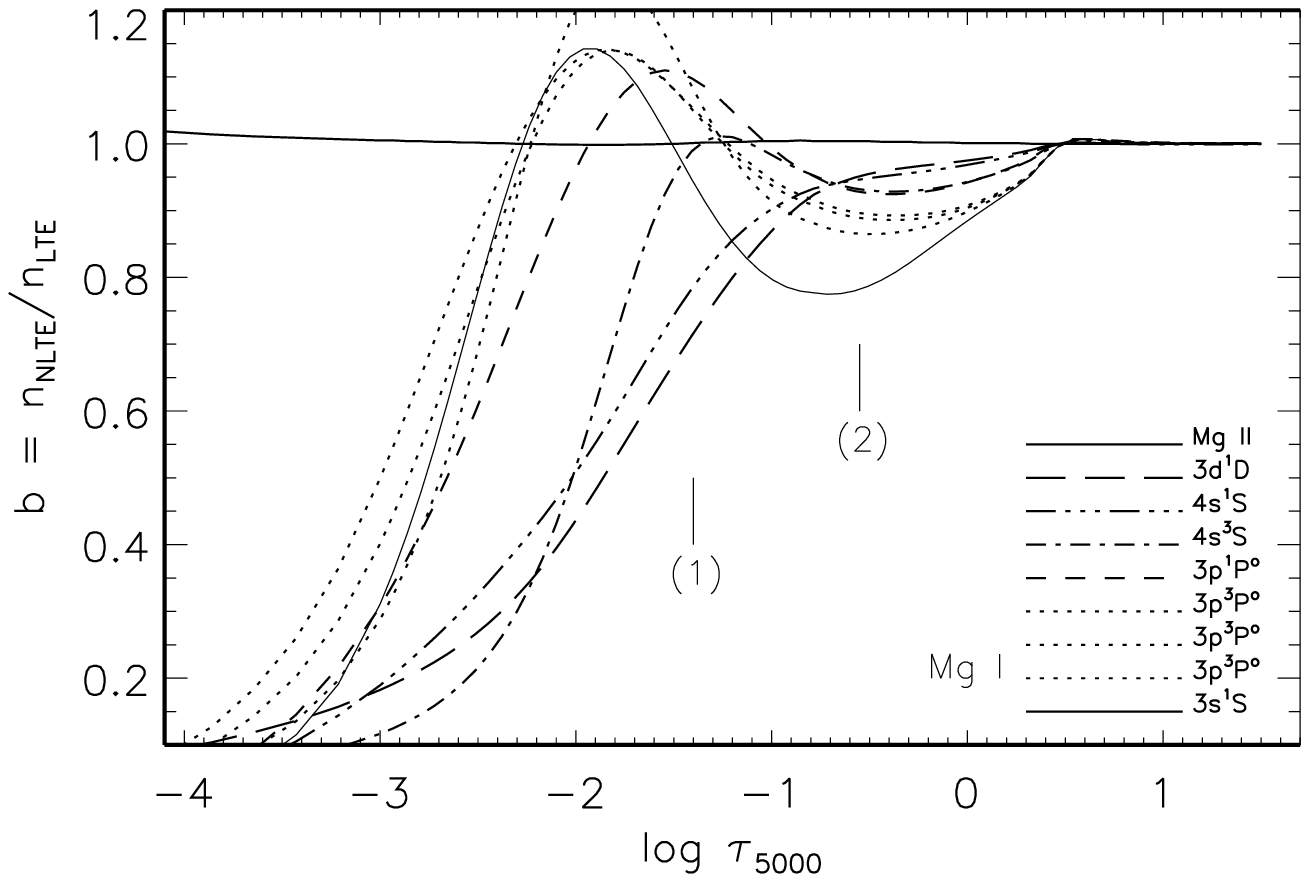}}
\caption{\label{bf_mg1} Departure coefficients for selected levels of
  \ion{Mg}{i} as a function of
  $\log \tau_{5000}$ in the model atmosphere 4600/1.60/$-2.56$ from the calculations with pure electronic collisions (top panel) and with \ion{H}{i} collisions taken into account following BBSGF (bottom panel). Tick marks
indicate the locations of line center optical depth unity for the \ion{Mg}{i} lines 5172\,\AA\ (1) and 5528\,\AA\ (2).}
\end{figure}

The departures from LTE for \ion{Mg}{i} grow toward lower metallicity due to decreasing number of electrons donated by metals and decreasing collision rates and also due to decreasing the UV opacity resulting in increasing photoionization rates. The largest uncertainties in NLTE results are caused by the uncertainties in collisional data. Here, we evaluate for the first time the effect of applying Mg+H collision data of \citet{mg_hyd2012} on the SE of magnesium in two well studied VMP stars. HD\,84937 represents the hot end of the stars that evolve on time scales comparable with the 
Galaxy lifetime. HD\,122563, in contrast, is a cool giant (Table~\ref{startab}).

Figure~\ref{bf_mg1} shows the departure coefficients $b_i = n_i^{\rm NLTE}/n_i^{\rm LTE}$ for the \ion{Mg}{i} levels in the model atmosphere 4600/1.60/$-2.56$ from calculations with two different treatment of collisional rates.
Here, $n_i^{\rm NLTE}$ and $n_i^{\rm LTE}$ are the SE and thermal (Saha-Boltzmann) number densities,
respectively. In the line-formation layers, the \ion{Mg}{i} levels have less populations compared with the thermal ones, 
independent of collision treatment, however, the magnitude of departures from LTE is significantly smaller in case of Mg+H collisions included compared with the case of pure electronic collisions. The main NLTE mechanism is overionization caused by superthermal radiation of a non-local origin below the thresholds of the \eu{3p}{3}{P}{\circ}{} and \eu{3p}{1}{P}{\circ}{} levels. When taking Mg+H collisions into account, 
the charge transfer processes \ion{Mg}{i} + \ion{H}{i} $\leftrightarrow$ \ion{Mg}{ii} + H$^-$ establish close collisional coupling of excited terms of \ion{Mg}{i} to the ground state of \ion{Mg}{ii} and reduce their underpopulation. This effect is redistributed to the lower excitation levels via the bound-bound transitions, including the \ion{H}{i} impact excitation and de-excitation processes.

\begin{table*}
 \centering
 \caption{\label{mg_cool} Magnesium LTE and NLTE abundances of HD\,84937 and HD\,122563 from calculations with different treatment of \ion{H}{i} collisions.}
  \begin{tabular}{cccccccccc}
   \hline\noalign{\smallskip}
$\lambda$ & \multicolumn{4}{c}{HD\,84937} & \ & \multicolumn{4}{c}{HD\,122563} \\
\noalign{\smallskip} 
\cline{2-5}
\cline{7-10} 
 (\AA) & LTE & pure e$^-$ & +H({\scriptsize BBSGF})$^1$ & +H({\scriptsize D0.1})$^2$ & & LTE & pure e$^-$ & +H({\scriptsize BBSGF}) & +H({\scriptsize D0.1}) \\
   \hline\noalign{\smallskip}
4571 & --6.38 & --6.24 & --6.30 & --6.33 & & --6.80 & --6.60 & --6.67 & --6.77 \\
4703 & --6.36 & --6.15 & --6.30 & --6.29 & & --6.86 & --6.59 & --6.82 & --6.84 \\
5528 & --6.34 & --6.15 & --6.30 & --6.27 & & --6.77 & --6.61 & --6.83 & --6.76 \\
5172 & --6.35$^3$ & --6.39 & --6.39 & --6.46 & & --6.87$^3$ & --6.76 & --6.95 & --6.96 \\
5183 & --6.34$^3$ & --6.39 & --6.39 & --6.46 & & --6.87$^3$ & --6.74 & --6.94 & --6.94 \\
\noalign{\smallskip} \hline \noalign{\smallskip}
 & \multicolumn{9}{c}{ \ \ \ $\Delta\log\varepsilon$(\ion{Mg}{i} 4703,5528 -- \ion{Mg}{i} 5172,5183)} \\
\noalign{\smallskip} \hline \noalign{\smallskip}
 &    --0.01  &   0.24 &   0.09 &   0.18 & &  0.05  &   0.15 &   0.12 &   0.15 \\
\noalign{\smallskip} \hline \noalign{\smallskip}
\multicolumn{10}{l}{Notes. \ $^1$ Mg+H collisions following BBSGF; 
 \ $^2$ Drawinian rates with \kH\,= 0.1; \ $^3$ from $EW$.} \\
\end{tabular}
\end{table*}

Table~\ref{mg_cool} presents abundances determined from individual lines of \ion{Mg}{i} under different assumptions for line formation. It is worth noting that results for \ion{Mg}{i} 4703\,\AA\ and 5528\,\AA\ do not depend on van der Waals damping treatment because both lines are weak. 
For each star, NLTE leads to depleted absorption in the lines at 4571, 4703, and 5528\,\AA\ due to overall overionization of \ion{Mg}{i} in the atmospheric layers, where they form. As expected, the inclusion of Mg+H collisions leads to weaker NLTE effects compared with that for pure electronic collisions, such that $\Delta_{\rm NLTE}$ reduces by 0.06-0.23~dex for different lines. For HD\,122563, the NLTE effects have different sign for \ion{Mg}{i} 4703\,\AA\ and \ion{Mg}{i} 5528\,\AA, resulting in a remarkable consistency of abundances derived from these two lines.

\begin{figure} 
  \resizebox{88mm}{!}{\includegraphics{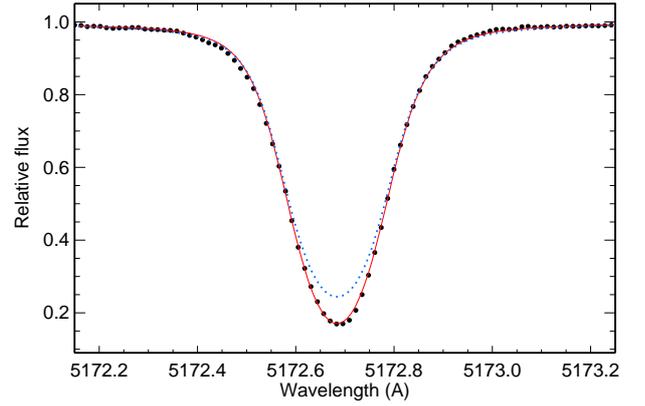}}
  \caption{\label{Fig:5172_hd122563} Best NLTE (continuous curve) and LTE (dotted curve) fits of the \ion{Mg}{i} 5172\,\AA\ line in HD\,122563 (bold dots). The NLTE calculations were performed with $\epsa{Mg} = -6.95$, while the LTE ones with a 0.02~dex higher value. Mg+H collisions are treated following BBSGF. 
Our theoretical flux profiles were convolved with a
profile that combines instrumental broadening with a Gaussian
profile and broadening by macroturbulence with a radial-tangential profile. 
The macroturbulence values are 4.7\,\kms\ and 3.8\,\kms\ for the NLTE and LTE fits, respectively.}
\end{figure}

Despite the low Mg abundance, the \ion{Mg}{i}b lines are rather strong in both VMP stars. For each line, its core forms in the layers, where departure coefficient of the upper level \eu{4s}{3}{S}{}{} drops rapidly (Fig.\,\ref{bf_mg1}) due to photon escape from the \ion{Mg}{i}b triplet lines themselves 
resulting in dropping the line source function below the Planck function and enhanced absorption in the line core (Fig.\,\ref{Fig:5172_hd122563}). NLTE leads to depleted absorption in the line wings due to overall overionization in deep atmospheric layers. However, this effect is predominated by NLTE strengthening for the line core, such that the obtained NLTE abundances are lower compared with the corresponding LTE ones and lower compared with those from the other lines, independent of collisional rate treatment. The difference $\Delta\log\varepsilon$(\ion{Mg}{i} 4703,5528 -- \ion{Mg}{i}b) is smaller in the case of Mg+H collisions included than in the case of pure electronic collisions and amounts to 0.09~dex and 0.12~dex for HD\,84937 and HD\,122563, respectively.
 For LTE abundances from the \ion{Mg}{i}b lines, we used $EW$s, because their profiles cannot be fitted at the LTE assumption. Despite the latter fact, the obtained LTE abundances are consistent within 0.05~dex with those from the other lines.


 Our calculations show that inelastic collisions with \ion{H}{i} atoms as treated by BBSGF produce significant thermalization effect on the SE of \ion{Mg}{i} in the atmospheres of VMP stars and improve Mg abundance determinations compared with the case of pure electronic collisions. To understand remaining discrepancies between different lines, one, probably, needs to go beyond the 1D analysis. The models based on hydrodynamical calculations (3D models) show, on average, lower temperatures outside $\log \tau_{5000} = -1$ compared with those in the corresponding 1D model, for example, by 1\,000~K at $\log \tau_{5000} = -2$ in the 4858/2.2/$-3$ model, according to \citet{Collet2007}. For VMP giants, these are exactly the layers where line core of the \ion{Mg}{i}b lines forms. 

\begin{table*}
 \centering
 \caption{\label{corr_cool} NLTE abundance corrections (dex) for lines of \ion{Mg}{i} from calculations with Mg+H collision rates of BBSGF and with the Drawinian rates scaled by a factor of \kH\,= 0.1 ({\scriptsize D0.1}).}
  \begin{tabular}{lrrcrrcrrcrrcrrcrr}
   \hline\noalign{\smallskip}
\multicolumn{1}{c}{Model} & \multicolumn{2}{c}{3829\,\AA} & & \multicolumn{2}{c}{5172\,\AA} & & \multicolumn{2}{c}{4571\,\AA} & & \multicolumn{2}{c}{4703\,\AA} & & \multicolumn{2}{c}{5528\,\AA} & & \multicolumn{2}{c}{5711\,\AA} \\
\noalign{\smallskip} \cline{2-3} \cline{5-6} \cline{8-9}\cline{11-12}\cline{14-15}\cline{17-18}
\noalign{\smallskip}
              & {\scriptsize BBSGF} &{\scriptsize D0.1} & &{\scriptsize BBSGF} &{\scriptsize D0.1} & &{\scriptsize BBSGF} &{\scriptsize D0.1} & &{\scriptsize BBSGF} &{\scriptsize D0.1} & &{\scriptsize BBSGF} &{\scriptsize D0.1} & &{\scriptsize BBSGF} &{\scriptsize D0.1} \\
\noalign{\smallskip} \hline \noalign{\smallskip}
6000/4.0/$-$1.0 & 0.05 & 0.02 & & 0.04 & 0.02    & & 0.08 & 0.05 & & 0.01 & 0.04 & & $-0.03$ & 0.04 & & 0.04 & 0.07 \\
6000/4.0/$-$2.0 & 0.04 & 0.01 & & 0.02 & $-0.04$ & & 0.08 & 0.04 & & 0.03 & 0.06 & & $-0.02$ & 0.06 & & 0.03 & 0.05 \\
6000/4.0/$-$3.0 & 0.08 & 0.02 & & 0.05 & $-0.04$ & &      &      & & 0.07 & 0.07 & &  0.07   & 0.08 & &      &  \\   
5811/4.0/$-$4.5 & 0.22 & 0.17 & & 0.17 & 0.19    & &      &      & &      &      & &         &      & &      & \\
6180/3.7/$-$4.0 & 0.25 & 0.19 & & 0.26 & 0.21    & &      &      & &      &      & &         &      & &      & \\
5000/2.0/$-$1.0 & 0.06 & 0.05 & & 0.06 & 0.05    & & 0.13 & 0.13 & & $-0.07$ & 0.00 & & $-0.16$ & $-0.07$ & & $-0.05$ &  0.06 \\
5000/2.0/$-$2.0 & 0.08 & 0.04 & & 0.03 & 0.00    & & 0.16 & 0.10 & & $-0.02$ & 0.07 & & $-0.12$ & 0.03 & & 0.03 & 0.09 \\
5000/2.0/$-$3.0 & 0.06 & 0.01 & & $-0.06$ & $-0.10$ & & 0.29 & 0.11 & & 0.18 & 0.16 & & 0.10 & 0.16 & & &  \\     
5100/2.2/$-$5.0 & 0.40 & 0.39 & & 0.34 & 0.37 & &  &    &  &   &   &  & &     & & & \\
\noalign{\smallskip} \hline \noalign{\smallskip}
\end{tabular}
\end{table*}

For most atoms, formalism of \citet{Drawin1968,Drawin1969} is still widely applied to simulate an additional source of thermalization in the atmospheres of cool stars by parametrized \ion{H}{i} collisions.
For \ion{Mg}{i}, \citet{mg_hyd2012} compared in their Fig.~1 the quantum scattering rate coefficients at $T$ = 6000~K with the results of the Drawin formula for five common bound-bound transitions. The Drawin/BBSGF rate ratio is close to unity for the \eu{3p}{3}{P}{\circ}{} - \eu{4s}{3}{S}{}{} transition, but the Drawinian rates are one to four orders of magnitude larger for the remianing transitions. However, relatively large rates compared to those
for excitation were obtained by BBSGF for the charge transfer processes. We find that they are larger compared to the Drawinian rates for ionization from the excited levels above \eu{3p}{1}{P}{\circ}{}, especially from the \eu{4s}{1}{S}{}{} state, where the difference is about four orders of magnitude at $T$ = 6000~K. It is worth noting that the Drawin/BBSGF rate ratios are only weakly sensitive to temperature variation. 

We employed the Drawinian rates scaled by a factor of \kH\ = 0.1 to determine magnesium abundances of the investigated stars (Table~\ref{mg_cool}). The changes in collisional rates result in different effects for 
different lines in different stars. For example, the use of accurate Mg+H collision data and scaled Drawinian rates leads to very similar results for \ion{Mg}{i} 4571, 4703, and 5528\,\AA\ in the model 6350/4.09/$-2.08$. For the cool giant model, the departures from LTE are larger when applying the BBSGF rates. However, nowhere the difference between using these two recipes exceeds 0.1~dex. This is smaller or similar to a line-to-line scatter for stellar abundance determinations including the present one.

The NLTE calculations were performed for a small grid of model atmospheres with hydrogenic collisions taken into account following BBSGF and \citet{Drawin1968,Drawin1969}. The obtained NLTE corrections are presented in Table\,\ref{corr_cool}. 
Their inspection will help the user to decide whether the Mg abundances determined earlier using the Drawinian rates should be revised. This depends on which spectral lines were employed in analysis and what was the range of stellar parameters. 

\section{Conclusions}\label{conclusion}

We summarize our findings.

1. For each of four investigated A-type stars, the use of $gf$-values of \citet{1990JQSRT..43..207C} provides consistent within 0.05~dex NLTE abundances from \ion{Mg}{i} 4703\,\AA, 5528\,\AA, and \ion{Mg}{i}b. The difference between different lines increases up to 0.11~dex, when employing experimental $gf$-values of \citet{Aldenius_mg1} for the \ion{Mg}{i}b lines.

2. We recommend to apply the corrections $\Delta\log\Gamma_6 = -0.3$ and $-0.2$ for \ion{Mg}{i} 4703\,\AA\ and 5528\,\AA, respectively, to the $ABO$ van der Waals damping constants, to remove the abundance discrepancy between different solar \ion{Mg}{i} lines. 

3. Inelastic collisions with \ion{H}{i} atoms as treated by \citet{mg_hyd2012} serve as efficient thermalizing process for the SE of \ion{Mg}{i} in the atmospheres of metal-poor stars. 
 The use of new collisional data improves Mg abundance determinations for HD\,84937 and HD\,122563, though does not remove completely the difference in abundance between different lines. 

\begin{acknowledgements}
L.\,M. thanks Paul Barklem for initiating this study and providing the $\Gamma_6$ value for \ion{Mg}{i} 4571\,\AA. 
 This study is supported by the Ministry of education and science of Russian Federation, project 8529 and 
the RF President with a grant on Leading Scientific Schools 3602.2012.2.  
\end{acknowledgements}

\bibliography{atomic_data,nlte,mp_stars}

\begin{thebibliography}{42}
\expandafter\ifx\csname natexlab\endcsname\relax\def\natexlab#1{#1}\fi

\bibitem[{{Aldenius} {et~al.}(2007){Aldenius}, {Tanner}, {Johansson},
  {Lundberg}, \& {Ryan}}]{Aldenius_mg1}
{Aldenius}, M., {Tanner}, J.~D., {Johansson}, S., {Lundberg}, H., \& {Ryan},
  S.~G. 2007, \aap, 461, 767

\bibitem[{{Allende Prieto} {et~al.}(2004){Allende Prieto}, {Asplund}, \&
  {Fabiani Bendicho}}]{h_cool_na_o}
{Allende Prieto}, C., {Asplund}, M., \& {Fabiani Bendicho}, P. 2004, \aap, 423,
  1109

\bibitem[{{Anstee} \& {O'Mara}(1995)}]{1995MNRAS.276..859A}
{Anstee}, S.~D. \& {O'Mara}, B.~J. 1995, \mnras, 276, 859

\bibitem[{{Asplund}(2005)}]{Asplund2005ARAA}
{Asplund}, M. 2005, \araa, 43, 481

\bibitem[{{Athay} \& {Canfield}(1969)}]{Athay1969mg}
{Athay}, R.~G. \& {Canfield}, R.~C. 1969, \apj, 156, 695

\bibitem[{{Barklem} {et~al.}(2011){Barklem}, {Belyaev}, {Guitou}, {Feautrier},
  {Gad{\'e}a}, \& {Spielfiedel}}]{Barklem2011_hyd}
{Barklem}, P.~S., {Belyaev}, A.~K., {Guitou}, M., {et~al.} 2011, \aap, 530, A94

\bibitem[{{Barklem} {et~al.}(2012){Barklem}, {Belyaev}, {Spielfiedel},
  {Guitou}, \& {Feautrier}}]{mg_hyd2012}
{Barklem}, P.~S., {Belyaev}, A.~K., {Spielfiedel}, A., {Guitou}, M., \&
  {Feautrier}, N. 2012, \aap, 541, A80

\bibitem[{{Barklem} \& {O'Mara}(1997)}]{1997MNRAS.290..102B}
{Barklem}, P.~S. \& {O'Mara}, B.~J. 1997, \mnras, 290, 102

\bibitem[{{Butler} \& {Giddings}(1985)}]{detail}
{Butler}, K. \& {Giddings}, J. 1985, Newsletter on the analysis of astronomical
  spectra, No. 9, University of London

\bibitem[{{Carlsson} {et~al.}(1992){Carlsson}, {Rutten}, \&
  {Shchukina}}]{Carlsson92}
{Carlsson}, M., {Rutten}, R.~J., \& {Shchukina}, N.~G. 1992, \aap, 253, 567

\bibitem[{{Chang} \& {Tang}(1990)}]{1990JQSRT..43..207C}
{Chang}, T.~N. \& {Tang}, X. 1990, \jqsrt, 43, 207

\bibitem[{{Collet} {et~al.}(2007){Collet}, {Asplund}, \&
  {Trampedach}}]{Collet2007}
{Collet}, R., {Asplund}, M., \& {Trampedach}, R. 2007, \aap, 469, 687

\bibitem[{{Dimitrijevic} \& {Sahal-Brechot}(1996)}]{mg1_g4}
{Dimitrijevic}, M.~S. \& {Sahal-Brechot}, S. 1996, \aaps, 117, 127

\bibitem[{{Drawin}(1968)}]{Drawin1968}
{Drawin}, H.-W. 1968, Zeitschrift fur Physik, 211, 404

\bibitem[{{Drawin}(1969)}]{Drawin1969}
{Drawin}, H.~W. 1969, Zeitschrift fur Physik, 225, 483

\bibitem[{{Fossati} {et~al.}(2009){Fossati}, {Ryabchikova}, {Bagnulo},
  {Alecian}, {Grunhut}, {Kochukhov}, \& {Wade}}]{2009A&A...503..945F}
{Fossati}, L., {Ryabchikova}, T., {Bagnulo}, S., {et~al.} 2009, \aap, 503, 945

\bibitem[{{Fossati} {et~al.}(2011){Fossati}, {Ryabchikova}, {Shulyak},
  {Haswell}, {Elmasli}, {Pandey}, {Barnes}, \& {Zwintz}}]{2011MNRAS.417..495F}
{Fossati}, L., {Ryabchikova}, T., {Shulyak}, D.~V., {et~al.} 2011, \mnras, 417,
  495

\bibitem[{Frebel {et~al.}(2005)Frebel, Aoki, Christlieb, Ando, Asplund,
  Barklem, Beers, Eriksson, Fechner, Fujimoto, Honda, Kajino, Minezaki, Nomoto,
  Norris, Ryan, Takada-Hidai, Tsangarides, \& Yoshii}]{Frebeletal:2005}
Frebel, A., Aoki, W., Christlieb, N., {et~al.} 2005, Nature, 434, 871

\bibitem[{{Gigas}(1988)}]{Gigas88}
{Gigas}, D. 1988, \aap, 192, 264

\bibitem[{{Gratton} {et~al.}(1999){Gratton}, {Carretta}, {Eriksson}, \&
  {Gustafsson}}]{Gratton1999}
{Gratton}, R.~G., {Carretta}, E., {Eriksson}, K., \& {Gustafsson}, B. 1999,
  \aap, 350, 955

\bibitem[{{Grupp} {et~al.}(2009){Grupp}, {Kurucz}, \& {Tan}}]{Grupp2009}
{Grupp}, F., {Kurucz}, R.~L., \& {Tan}, K. 2009, \aap, 503, 177

\bibitem[{{Idiart} \& {Th{\'e}venin}(2000)}]{Thevenin2000}
{Idiart}, T. \& {Th{\'e}venin}, F. 2000, \apj, 541, 207

\bibitem[{{Kupka} {et~al.}(1999){Kupka}, {Piskunov}, {Ryabchikova}, {Stempels},
  \& {Weiss}}]{vald}
{Kupka}, F., {Piskunov}, N., {Ryabchikova}, T.~A., {Stempels}, H.~C., \&
  {Weiss}, W.~W. 1999, \aaps, 138, 119

\bibitem[{{Kurucz} {et~al.}(1984){Kurucz}, {Furenlid}, {Brault}, \&
  {Testerman}}]{Atlas}
{Kurucz}, R.~L., {Furenlid}, I., {Brault}, J., \& {Testerman}, L. 1984, {Solar
  flux atlas from 296 to 1300 nm} (New Mexico: National Solar Observatory)

\bibitem[{{Lemke} \& {Holweger}(1987)}]{Lemke1987}
{Lemke}, M. \& {Holweger}, H. 1987, \aap, 173, 375

\bibitem[{{Lodders} {et~al.}(2009){Lodders}, {Plame}, \& {Gail}}]{Lodders2009}
{Lodders}, K., {Plame}, H., \& {Gail}, H.-P. 2009, in Landolt-B{\"o}rnstein -
  Group VI Astronomy and Astrophysics Numerical Data and Functional
  Relationships in Science and Technology Volume 4B: Solar System. Edited by
  J.E. Tr{\"u}mper, 2009, 4.4., 44--54

\bibitem[{{Mashonkina} {et~al.}(2011){Mashonkina}, {Gehren}, {Shi}, {Korn}, \&
  {Grupp}}]{mash_fe}
{Mashonkina}, L., {Gehren}, T., {Shi}, J.-R., {Korn}, A.~J., \& {Grupp}, F.
  2011, \aap, 528, A87

\bibitem[{{Mashonkina} {et~al.}(2007){Mashonkina}, {Korn}, \&
  {Przybilla}}]{mash_ca}
{Mashonkina}, L., {Korn}, A.~J., \& {Przybilla}, N. 2007, \aap, 461, 261

\bibitem[{{Mashonkina} {et~al.}(1996){Mashonkina}, {Shimanskaya}, \&
  {Sakhibullin}}]{1996ARep...40..187M}
{Mashonkina}, L.~I., {Shimanskaya}, N.~N., \& {Sakhibullin}, N.~A. 1996,
  Astronomy Reports, 40, 187

\bibitem[{{Mauas} {et~al.}(1988){Mauas}, {Avrett}, \& {Loeser}}]{Mauas1988}
{Mauas}, P.~J., {Avrett}, E.~H., \& {Loeser}, R. 1988, \apj, 330, 1008

\bibitem[{{Merle} {et~al.}(2011){Merle}, {Th{\'e}venin}, {Pichon}, \&
  {Bigot}}]{2011MNRAS.418..863M}
{Merle}, T., {Th{\'e}venin}, F., {Pichon}, B., \& {Bigot}, L. 2011, \mnras,
  418, 863

\bibitem[{{Mishenina} {et~al.}(2004){Mishenina}, {Soubiran}, {Kovtyukh}, \&
  {Korotin}}]{Mishenina2004}
{Mishenina}, T.~V., {Soubiran}, C., {Kovtyukh}, V.~V., \& {Korotin}, S.~A.
  2004, \aap, 418, 551

\bibitem[{{Przybilla} {et~al.}(2001){Przybilla}, {Butler}, {Becker}, \&
  {Kudritzki}}]{Przybilla_mg}
{Przybilla}, N., {Butler}, K., {Becker}, S.~R., \& {Kudritzki}, R.~P. 2001,
  \aap, 369, 1009

\bibitem[{{Przybilla} {et~al.}(2011){Przybilla}, {Nieva}, \&
  {Butler}}]{2011JPhCS.328a2015P}
{Przybilla}, N., {Nieva}, M.-F., \& {Butler}, K. 2011, Journal of Physics
  Conference Series, 328, 012015

\bibitem[{{Ralchenko} {et~al.}(2008){Ralchenko}, {Kramida}, {Reader}, \&
  Team}]{NIST}
{Ralchenko}, Y.~A., {Kramida}, E., {Reader}, J., \& Team, N.~A. 2008, NIST
  Atomic Spectra Database (version 3.1.5) (USA)

\bibitem[{Reetz(1991)}]{Reetz}
Reetz, J.~K. 1991, Diploma Thesis (Universit\"at M\"unchen)

\bibitem[{{Seaton}(1962)}]{Seaton1962}
{Seaton}, M.~J. 1962, Atomic and Molecular Processes(New York: Academic Press)
  (New York: Academic Press)

\bibitem[{{Shulyak} {et~al.}(2004){Shulyak}, {Tsymbal}, {Ryabchikova},
  {St{\"u}tz}, \& {Weiss}}]{LLMODELS}
{Shulyak}, D., {Tsymbal}, V., {Ryabchikova}, T., {St{\"u}tz}, C., \& {Weiss},
  W.~W. 2004, \aap, 428, 993

\bibitem[{{Sitnova} {et~al.}(2013){Sitnova}, {Mashonkina}, \&
  {Ryabchikova}}]{sitnova_o}
{Sitnova}, T.~M., {Mashonkina}, L.~I., \& {Ryabchikova}, T.~A. 2013, Astronomy
  Letters, 38, in press

\bibitem[{{Steenbock} \& {Holweger}(1984)}]{Steenbock1984}
{Steenbock}, W. \& {Holweger}, H. 1984, \aap, 130, 319

\bibitem[{{Tolstoy} {et~al.}(2009){Tolstoy}, {Hill}, \&
  {Tosi}}]{2009ARA&A..47..371T}
{Tolstoy}, E., {Hill}, V., \& {Tosi}, M. 2009, \araa, 47, 371

\bibitem[{{Zhao} {et~al.}(1998){Zhao}, {Butler}, \& {Gehren}}]{Zhao1998}
{Zhao}, G., {Butler}, K., \& {Gehren}, T. 1998, \aap, 333, 219

\end{thebibliography}
\bibliographystyle{aa}


\end{document}